\documentclass[twocolumn,amsmath,amssymb,amsfonts,superscriptaddress
,floatfix,showpacs,prd,aps]{revtex4}
\newcommand{\del}{\partial}
\newcommand{\Tr}{\mathrm{Tr}}

\newcommand{\dslash}{\partial\!\!\!/}

\def\sl#1{\setbox0=\hbox{$#1$}               % set a box for #1
   \dimen0=\wd0                                 % and get its size
   \setbox1=\hbox{/} \dimen1=\wd1               % get size of /
   \ifdim\dimen0>\dimen1                        % #1 is bigger
      \rlap{\hbox to \dimen0{\hfil/\hfil}}      % so center / in box
      #1                                        % and print #1
   \else                                        % / is bigger
      \rlap{\hbox to \dimen1{\hfil$#1$\hfil}}   % so center #1
      /                                         % and print /
   \fi}                                         %
\usepackage{graphicx,color,amsmath}

\begin{document}
\title{\bf The Functional Renormalization Group and $O(4)$ scaling}
\author{B.~Stoki\'c}
\affiliation{%
Gesellschaft f\"ur Schwerionenforschung, GSI,  D-64291 Darmstadt, Germany}
\author{B.~Friman}
\affiliation{%
Gesellschaft f\"ur Schwerionenforschung, GSI,  D-64291 Darmstadt, Germany}
\author{K.~Redlich}
\affiliation{%
Institute of Theoretical Physics, University of Wroclaw, PL--50204 Wroc\l aw,
Poland}
\affiliation{%
Institut f\"ur Kernphysik, Technical University Darmstadt, D-64289 Darmstadt,
Germany}
%\email[E-Mail:]{}
%\affiliation{}


\date{\today}

\begin{abstract}
The critical behavior of the chiral quark-meson model is studied within the
Functional Renormalization Group (FRG). We derive the flow equation for the
scale dependent thermodynamic potential at finite temperature and density in the
presence of a symmetry-breaking external field. Within this scheme, the critical
scaling behavior of the order parameter, its transverse and longitudinal
susceptibilities as well as the correlation lengths near the chiral phase
transition are computed. We focus on the scaling properties of these observables
 at non-vanishing   external  field when  approaching  the critical point from
the symmetric as well as from the broken phase. We confront our numerical
results with the Widom-Griffiths form of the magnetic equation of state,
obtained by a systematic $\epsilon$--expansion of the scaling function. Our
results for the critical exponents are consistent with those recently computed
within Lattice Monte-Carlo studies of the $O(4)$ spin system.

\end{abstract}

\maketitle
%%%%%%%%%%%%%%%%%%%%%%%%%%%%%%%
%  SECTION I introduction
%%%%%%%%%%%%%%%%%%%%%%%%%%%%%%%

\section{Introduction}

Understanding the phase structure and the critical properties of strongly
interacting matter is one of the central problems addressed in studies of QCD
thermodynamics. Lattice Gauge Theory (LGT) calculations at finite temperature
show a clear separation between the confined (hadronic) and deconfined
(quark-gluon plasma) phases \cite{LGT}.

Universality arguments imply that two-flavor QCD exhibits
a second order chiral phase transition belonging to the same universality class
as the $O(4)$ spin system in three dimensions ~\cite{Pisarski:1983ms,
Rajagopal:1992qz,
ParisenToldin:2003hq,JE1,JE2,Engels:2003nq,fk}. Consequently, the long-range
properties  of the chiral phase transition  can be explored within effective
models of the same universality class, independent of the specific dynamics.

One such model is the chiral quark-meson model, which often is used as an
effective realization of the low-energy
sector of QCD \cite{Schaefer:npa}.
Thus, by studying the  thermodynamics of this model near the phase transition,
one can explore the leading singularities of thermodynamic quantities in
two-flavor QCD at the second order chiral phase transition. In other words, we
can find
the universal properties of dynamical chiral symmetry breaking at
finite temperature and/or density. However, at this point we stress that a
description of
the QCD phase transition within effective models is subject to limitations.
First, in chiral effective models one cannot address the deconfinement
phenomenon in QCD and second, one cannot
obtain quantitative information on the thermodynamics outside the chiral
critical region. More generally, non-universal quantities depend on the specific
dynamics of the theory and can therefore not be computed reliably within
effective models.

As a rule, phase transitions and critical phenomena are related with singular
behavior of some susceptibility which in turn provide a measure of the
fluctuations of a physical
quantity.  At a  second order phase transition, there are  long-range
critical correlations that appear due to the presence of massless modes. The
long-range correlations imply divergent susceptibilities and consequently large
fluctuations.

In the mean field approximation, which is often used to describe the chiral
phase transition
\cite{weise,JW,CS}, one fails to obtain a quantitatively correct description of
critical phenomena,
since the influence of fluctuations and non-perturbative effects
near the phase transition are neglected. However, methods based on the
renormalization group (RG)
can account for both of these important effects  (see e.g.~\cite{Ma}). The
advantage of the
RG method is that it  can describe physics across different momentum scales. In
particular, within the RG framework one can capture the dynamics of the
long-range
fluctuations near the critical point. Wilsonian RG techniques, such as the
functional
renormalization group (FRG), are particularly  useful in describing phase
transitions
\cite{Berges:review}. One of the important results of RG theory is that it
provides a rationale for universality, i.e., the fact that models can be grouped
into
universality classes that, depending only on the dimension of the system and the
symmetry of the order parameter, show the same critical behavior.

In this paper we use the FRG method in the quark--meson model to explore the
universal
properties of the chiral phase transition. This model has
been studied using different RG approaches~\cite{Jungnickel, Berges:epjc,
Meyer:vac,
Braun:2003ii,Schaefer:npa} both in vacuum as well as in medium, at
finite
temperature and chemical potential. Various  critical exponents
have also been computed~\cite{Berges:prd_qm, Tetradis1, Schaefer:vac,
Braun:Klein}.
In this paper we go beyond previous RG studies of this model. We analyze the
critical region in the
broken and in the symmetric phase close to the critical point and
discuss the scaling functions for different physical quantities as well as the
magnetic equation
of state \cite{Onuki}.   Furthermore, we compute
the critical exponents for the order parameter and its transverse and
longitudinal
susceptibilities as well as for the correlation lengths. We also calculate the
effective critical exponents~\cite{Wegner} in the presence
of the chiral symmetry breaking field.
Finally, we
explore the scaling behavior of different ratios of susceptibilities and  the
order parameter \cite{Boyd}.

We also show that within the FRG method, with a suitable truncation  of
the FRG flow equations, the quark-meson model exhibits a universal behavior
that is in quantitative agreement  with that recently found in  lattice
calculations  of the $O(4)$ spin system in three
dimensions~\cite{Kanaya:1994qe}.

The paper is organized as follows: In Sec.~\ref{sec:frg} we introduce  the
chiral
quark-meson model and the FRG method. The flow equations for the scale
dependent
thermodynamic potential at finite temperature and chemical potential are also
derived
in this Section. In Sec.~\ref{sec:crit} we discuss the universal scaling
properties of
various physical quantities and extract their critical exponents. In
Sec.~\ref{sec:concl} we  summarize our results and discuss  their relevance.

\section{The chiral quark-meson model and the FRG flow equation}\label{sec:frg}

We employ the chiral quark--meson model to explore the scaling properties and
the critical
equation of state within the FRG method. This model is relevant for studying
strongly interacting hot and dense matter with two degenerate light--quark
flavors, since it is expected  to belong to the same universality class as QCD
at the chiral phase transition.

The chiral quark-meson model represents an effective low-energy realization
for spontaneous chiral symmetry breaking at the intermediate momentum scale
$4\pi f_\pi\approx 1$ GeV. One can view this model as an effective model of
QCD, where the gluon degrees of freedom have been integrated out. Consequently,
the model does not describe the confinement-deconfinement transition.

The Lagrangian density of the chiral quark-meson model is given by
\begin{equation}\label{eq:qmmodel}
  {\cal L}=\frac{1}{2}\left(\del_{\mu}\phi\right)^2+ \bar{q}i\sl{\del}q+
g\bar{q}M q+U(\sigma,\vec{\pi}),
\end{equation}
where the $O(4)$ representation of the meson fields is
$\phi=(\sigma,\vec{\pi})$ and the
corresponding $SU(2)_L\otimes SU(2)_R$ chiral  representation is given by
$M=\sigma+i\vec{\tau}\cdot\vec{\pi}\gamma_5$. There are $N_f^2=4$
mesonic
degrees of freedom coupled to $N_f=2$ quark flavors $q$.

The potential $U(\sigma,\vec{\pi})$ is  given by
\begin{equation}
U(\sigma,\vec{\pi})=\frac{1}{2}m^2\phi^2+\frac{\lambda}{4}\phi^4-c\sigma.
\label{pot}
\end{equation}
In vacuum, the $O(4)$ symmetry of the Lagrangian (\ref{eq:qmmodel}) is,  for
$m^2<0$, spontaneously broken
to $O(3)$. This leads to a nonvanishing scalar condensate \footnote{The scalar
condensate is proportional to the quark condensate
$\langle\bar{q} q\rangle$.}
$\langle\sigma\rangle=f_{\pi}$. The explicit symmetry breaking term $c\sigma$
in the potential gives the pion a mass $m_\pi$.  At the three level,
$c_0=f_{\pi}m_{\pi}^2$ yields the physical pion mass in vacuum. For convenience
we introduce a dimensionless parameter $h=c/c_0$, the reduced external field, as
a measure for the strength of the symmetry breaking term. In a medium, the
chiral symmetry of the Lagrangian is restored, leading to a vanishing  scalar
condensate at  some critical temperature and/or density.

\begin{figure}[t!]
\begin{center}
\includegraphics*[width=8.6cm]{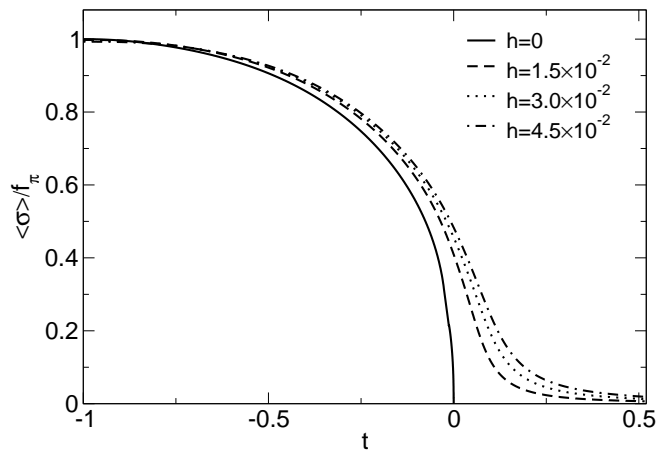}
\caption{The order parameter as a function of the reduced temperature $t$ for
several values of the external field $h$.} \label{magnet}
\end{center}
\end{figure}

%%%%%%%%% FRG part %%%%%%%%%%%%%%%%%%%%%%%%%%%%%%%%%%%%%%%%%%%%%%%%%%%%%%%%%
To explore the critical properties near the chiral phase transition we use the
Functional Renormalization Group. The FRG  is an important
tool for addressing nonperturbative problems within quantum field theory. The
starting point is the infrared (IR) regularized effective average action
$\Gamma_k$~\cite{Wetterich, Morris, Ellwanger, Berges:review}. This is a
generalization of the standard effective
action, where one-particle modes with a momentum less than a scale parameter
$k$ are suppressed by means of a regulator function.

The FRG equation for the effective action reads~\cite{Wetterich, Berges:review}
\begin{eqnarray}\label{ERG}
\del_k\Gamma_k[\Phi,\psi]&=&\frac{1}{2}\Tr\left\{\del_k
R_{kB}\biggl(\Gamma_k^{(2)}
[\Phi,\psi]+R_{kB}\biggr)^{-1}\right\}\nonumber \\
&&\hspace*{-7mm}-\Tr\left\{\del_k
R_{kF}\left(\Gamma_k^{(2)}[\Phi,\psi]+R_{kF}\right)^{-1}\right\},
\end{eqnarray}
where $\Gamma_k^{(2)}$ denotes the second functional derivative of
$\Gamma_k[\Phi,\psi]$
with respect to the field variables and corresponds
to the inverse two-point function at the scale $k$.
The trace in Eq. (\ref{ERG}) denotes  a momentum integration and a summation
over all internal indices like e.g. flavor, color, and/or Dirac indices.
The fields $\Phi$ and $\psi$ denote bosonic and fermionic fields, respectively.

The effective average action $\Gamma_k$ governs the dynamics of a theory at the
momentum
scale $k$ and interpolates between the bare action $\Gamma_{k=\Lambda}\equiv S$
and the full
quantum effective action $\Gamma_{k=0}=\Gamma$. The regulator function $R_k$,
which suppresses the small momentum modes, is to some extent arbitrary. The
derivative
of this function $\partial_tR_k$ is peaked at the scale $k$ and thus implements
the Wilsonian idea of successively integrating out shells in momentum space.

The effect of fluctuations is thus gradually included in $\Gamma_k$ by solving
the FRG flow equation. In Euclidean space-time
\begin{eqnarray}\label{eq:trunc}
\Gamma_k &=& \int d^4x\left\{\frac{1}{2}Z_{\phi,k}\left(\del_{\mu}\phi\right)^2
+
Z_{\psi,k}\bar{\psi}\dslash\psi   \right.\nonumber \\
&& +\left.
g\bar{\psi} \, (\sigma + i
  \vec \tau\cdot \vec \pi\bar{\gamma} )\,\psi + U_k(\rho,\sigma)
  \right\},
\end{eqnarray}
where the field $\rho$ is given by
\begin{equation}\label{eq5}
\rho=\frac{1}{2}\phi^2=\frac{1}{2}\left(\sigma^2+\vec{\pi}^2 \right)
\end{equation}
and the fermionic field $\psi$ carries  two flavors corresponding to the up and
down
quarks. The  matrix $\bar{\gamma}$ is the Euclidean analogue of $\gamma_5$.

In the following  we assume that each spectral function is dominated by a pole,
corresponding to a quasiparticle. Furthermore, we neglect the wavefunction
renormalization for both the bosonic
and fermionic fields ($Z_{\phi,k}=Z_{\psi,k}=1$), i.e. the anomalous dimension
is set to
zero. Thus, changes of quasiparticle properties in the medium are accounted
for, but the fragmentation of singel-particle strength is ignored. Finally, we
neglect the scale dependence of the Yukawa coupling $g$ in Eq.
(\ref{eq:trunc}). Consequently, the only scale dependence we are left with is
that of the potential $U_k(\rho)$.

This approximation to the average effective action   $\Gamma_k$ in
(\ref{eq:trunc}) corresponds to the leading order derivative expansion, which is
obtained as the leading term in a
systematic expansion in powers of  derivatives of the fields. For uniform field
configurations, the average effective action $\Gamma_k$ evaluated in this
approximation is
proportional to the effective potential $U_k(\rho,\sigma)$, since
\begin{equation}
\Gamma_k=\int d^4x\, U_k(\rho,\sigma).
\end{equation}
The leading order derivative expansion offers a relatively transparent
framework,
which nonetheless yields a very useful description of the critical fluctuations
near a phase transition.

In addition to defining the form of the effective action, one also needs to
specify the
regulator functions $R_k$. If
suitably chosen, the momentum integration is both infrared and ultraviolet
finite, leading
to a numerically stable solution of the flow equations. We use the so called
optimized regulator functions~\cite{Litim:opt},
\begin{equation}\label{eq:BRb}
R^{\mathrm{opt}}_{B,k}(q^2)=(k^2-q^2)\theta(k^2-q^2),
\end{equation}
for bosons and
\begin{equation}\label{eq:BRf}
R^{\mathrm{opt}}_{F,k}(q)=
\sl{q}
\left(\sqrt\frac{k^2}{q^2}-1\right)\theta(k^2-q^2)
\end{equation}
for fermions. Such regulator functions in general lead to a stabilization of the
RG flow and an improved convergence towards the physical
theory~\cite{Litim:opt}. Due to the theta function, the regulator acts as a
momentum-dependent mass term for all modes with
$q^2\leq k^2$.

%%%%%%%%%%%%%%%%%%%%%%%%%%%%%%%
% SECTION II, part A
%%%%%%%%%%%%%%%%%%%%%%%%%%%%%%%

\subsection{Flow at  finite temperature  and chemical potential}

In the previous section we have introduced the basic concepts of FRG. In the
following we
will use this method to find the nonperturbative thermodynamic potential at
finite
temperature and density.

We treat bosons and fermions in thermal equilibrium in the
standard imaginary time formalism. In thermal equilibrium  the (fermion)
boson fields
satisfy  (anti-) periodic  boundary conditions in the Euclidean time direction
with periodicity $1/T$. The momentum integration is replaced by a Matsubara
sum as follows:
\begin{equation}
\int\frac{d^dq}{(2\pi)^d}\rightarrow
T\sum_{n\in\mathbb{Z}}\int\frac{d^{d-1}q}{(2\pi)^{d-1}}
\end{equation}
where
\begin{equation}
q_0(n)=2n\pi T, \;\qquad q_0(n)=(2n+1)\pi T
\end{equation}
for bosons and fermions, respectively.

We employ a modified form of the optimized regulator, where the Euclidean
4-momentum squared $q^2$ in Eqs. (\ref{eq:BRb}) and (\ref{eq:BRf}) is replaced
by the 3-momentum squared $\mathbf{q}^2$. Thus, for bosons we use~\cite{Blaizot}
\begin{equation}\label{eq:BRb2}
R^{\mathrm{opt}}_{B,k}(\mathbf{q})=(k^2-\mathbf{q}^2)\theta(k^2-\mathbf{q}^2).
\end{equation}
The finite quark chemical potential $\mu$ is introduced by the  following
substitution of the time derivative
\begin{equation}
\del_0\rightarrow \del_0+i\mu
\end{equation}
in the fermionic part of the effective action (\ref{eq:trunc}). At finite
$\mu$, the fermionic regulator function is then of the form
\begin{equation}\label{eq:BRf2}
R^{\mathrm{opt}}_{F,k}(q)= \left(\sl{q}+i\mu\gamma^0\right)
\left(\sqrt\frac{(q_0+i\mu)^2+k^2}{(q_0+i\mu)^2+\mathbf{q}^2}
-1\right)\theta(k^2-\mathbf{q} ^2).
\end{equation}
Clearly, the modified regulators are invariant only under spatial rotations of
the momentum $\mathbf{q}$ in a particular frame, but not under Euclidean
rotations involving the imaginary time direction. However, this is not a crucial
issue in calculations at finite $T$ and $\mu$, where the heat bath defines a
preferred frame. Furthermore, the Euclidean invariant form leads to various
difficulties, as pointed out in \cite{Blaizot}. A great advantage of the
regulators (\ref{eq:BRb2}) and (\ref{eq:BRf2}) is that, in the quasiparticle
approximation, the Matsubara sums for the one-loop diagrams are identical to
those appearing in a free theory.

\begin{figure*}
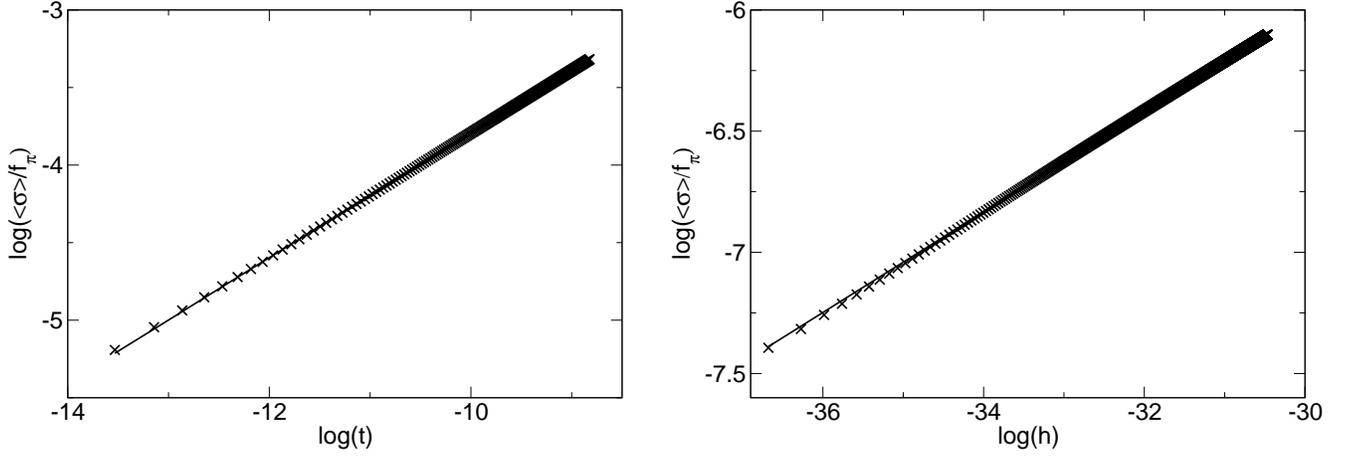

\begin{center}
\includegraphics*[height=6cm]{log_beta.eps}\hspace{5mm}
\includegraphics*[height=6cm]{log_delta.eps}
\caption{Scaling of the order parameter with $t$ and $h$. The logarithmic fits
yield the critical exponents $\beta$ (left panel) and $\delta$ (right panel).}
\label{log_beta_delta}
\end{center}
\end{figure*}

Thus, with the regulators (\ref{eq:BRb2}) and (\ref{eq:BRf2}) one obtains the
following flow equation for the thermodynamic potential density
($\Omega=(T/V)\Gamma$)
\begin{widetext}
\begin{equation}\label{eq:frg_flow}
\del_k \Omega_k(T,\mu)=\frac{k^4}{12\pi^2}
 \left[\frac{3}{E_\pi} \Bigg( 1+2n_B(E_\pi)\Bigg)
 +\frac{1}{E_\sigma} \Bigg( 1+2n_B(E_\sigma)
    \Bigg) -\frac{2\nu_q}{E_q} \Bigg(
    1-n_F(E_q)-\bar n_F(E_q) \Bigg) \right],
\end{equation}
\end{widetext}
where $n_B(E_{\pi,\sigma})$ and $n_F(E_q),\bar{n}_F(E_q) $ are the bosonic and
fermionic distribution functions
\begin{equation*}
n_B(E_{\pi,\sigma})=\frac{1}{e^{E_{\pi,\sigma}/T}-1},
\end{equation*}
\begin{equation*}
n_F(E_q)=\frac{1}{e^{
    (E_q-\mu)/T}+1}  , \quad \bar n_F(E_q)=\frac{1}{e^{(E_q+\mu)/T}+1}.
\end{equation*}
Here
\begin{equation*}
E_\pi = \sqrt{k^2+\bar{\Omega}'_k}\;~,~ E_\sigma =\sqrt{k^2+
\bar{\Omega}'_k+2\rho \bar{\Omega}''_k},
\end{equation*}
is the  pion and sigma energies, where $\bar{\Omega}_k'$ and
$\bar{\Omega}_k{''}$ denote the first and the second derivatives of
the thermodynamic potential with the explicit symmetry breaking term removed,
$\bar{\Omega}_k=\Omega_k+c\sqrt{2\,\rho_k}$, with respect to $\rho$ and
\begin{equation*}
E_q =\sqrt{k^2+2g^2\rho}
\end{equation*}
the  quark energy. Finally, the quark degeneracy is given by $\nu_q=2N_cN_f$.

The flow equation (\ref{eq:frg_flow}) is identical to the one obtained within
the proper time renormalization group  (PTRG) scheme~\cite{Braun:2003ii,
Schaefer:npa}. In fact, it was shown by Litim~\cite{Litim:opt}
that, in vacuum, the PTRG scheme, with a properly chosen cutoff function, is
equivalent to the optimized FRG flow in the leading order derivative expansion.

The flow equation for  $\Omega_k(T,\mu)$ can be solved numerically by either
discretizing the potential on a grid~\cite{Schaefer:npa} or expanding it in
powers of the
fields $(\sigma,\rho)$. In this paper, we employ the second method and expand
the potential
$\Omega_k(T,\mu)$ in a Taylor series
around the minimum at $\sigma_0=\sqrt{2\rho_0}$
\begin{equation}\label{eq:taylor}
\bar{\Omega}_k(T)=\sum_m \frac{a_{m,k}(T)}{m!}(\rho_k-\rho_0)^m.
\end{equation}
The symmetry breaking term is treated explicitly, leading to
\begin{equation}\label{eq:symm-break}
\Omega_k(T)=\bar{\Omega}_k(T)-c\sigma_k.
\end{equation}
The stationarity condition
\begin{equation}\label{eq:min}
\frac{\del \Omega_k(T)}{\del\sigma}\vert_{\mathrm{min}}=0,
\end{equation}
determines the position of the physical minimum at the scale $k$. Thus, for
$c\neq 0$, i.e., for an explicitly broken symmetry, we find
\begin{equation}
c=a_1\sigma_0
\end{equation}
which relates the coupling $a_1$ and the expectation value of the scalar field
$\sigma_0$.

We truncate the expansion in Eq.~(\ref{eq:taylor}) at $m=3$. Using the
identity~\footnote{We note that $\del\Omega_k/\del \rho$ vanishes, due to the
gap equation (\ref{eq:min}), so that $d\Omega_k/d k=\del\Omega_k/\del k$.}
\begin{equation}\label{eq:identity}
\frac{d \Omega'_k}{d k}=\frac{\del\Omega'_k}{\del\rho}\frac{d\rho}{d k}
+\frac{\del
\Omega'_k}{\del k}\, ,
\end{equation}
and the corresponding relations for $\Omega''_k$ and $\Omega'''_k$, we then
obtain the flow equations
\begin{subequations}\label{eq:rgflow}
\begin{eqnarray}
d_ka_0&=&\frac{c}{\sqrt{2\rho_k}}\,d_k\rho_k+\del_k\Omega_k,
\\
d_k\rho_k&=&-\frac{1}{\left(c/(2\rho_k)^{3/2}+a_2\right)}
\del_k\Omega'_k,\label{eq:trunc2}
\\
d_ka_2&=&a_3\, d_k\rho_k+\del_k\Omega''_k,
\\
d_ka_3&=&\del_k\Omega'''_k,\label{eq:trunc4}
\end{eqnarray}
\end{subequations}
where we have introduced the shorthand notation $d_k=d/dk$.
In Eq.~(\ref{eq:identity}) the terms $\del\Omega'_k/\del\rho$ etc. are
evaluated using the expansion (\ref{eq:taylor}) truncated at $m=3$.

In the following we consider only the case of vanishing net baryon density,
i.e., $\mu=0$, while the external field $c$ is kept finite. Results for the
chiral limit are then obtained by letting $c\rightarrow0$.

The flow equations are solved numerically starting at a cutoff scale
$\Lambda=1.2$ GeV. The initial values for the parameters are chosen  such that
the physical pion mass $m_{\pi}=138$ MeV, the pion decay constant $f_{\pi}=93$
MeV and the sigma mass $m_{\sigma}=600$ MeV are reproduced in vacuum, using the
regulators (\ref{eq:BRb2}) and (\ref{eq:BRf2}), for a constituent quark mass of
$m_q=300$ MeV. The strength of the
external field $c_0$ and the Yukawa coupling ($g=3.2$) are fixed by the pion
mass and the constituent quark mass. The pion and  the sigma
masses, at a given momentum scale $k$, are given by
\begin{equation}
m_{\pi,k}^2=\frac{c}{\sqrt{2\rho_k}},\quad
m_{\sigma,k}^2=\frac{c}{\sqrt{2\rho_k}}+2\rho_k a_{2,k}\, ,
\end{equation}
while  the constituent quark mass is obtained from
\begin{equation}
m_{q,k}^2=2g^2\rho_k.
\end{equation}

The numerical solution of the flow equations yields a non-perturbative
thermodynamic potential at finite temperatures. This can then be used to
explore the critical properties of thermodynamical observables in the
vicinity of the chiral phase transition.

%%%%%%%%%%%%%%%%%%%%%%%%%%%%%%%%%%%%%%
% SECTION III
%%%%%%%%%%%%%%%%%%%%%%%%%%%%%%%%%%%%%%

\section{Critical behavior and $O(4)$ scaling}\label{sec:crit}

It is well known that the effect of critical fluctuations on the properties of
a system close to a second-order phase transition can be efficiently computed by
means of the Wilsonian renormalization group~\cite{wilson-kogut}. A systematic
account of fluctuations is essential for a realistic description of the critical
properties of such systems. The FRG approach offers a powerful tool for
computing not only universal quantities, like the critical exponents, but also
the full thermodynamics of a strongly interacting many-body systems. Here we
focus on the critical exponents of the chiral phase transition.

In the chiral
limit, the chiral quark-meson model exhibits a second order phase transition.
The critical behavior of this model (and of QCD) is expected to be governed by
the fixed point of the $O(4)$ Heisenberg model in three dimensions. Universality
class arguments lead to predictions for the functional form of various
thermodynamic quantities in the vicinity of the critical temperature $T_c$, the
critical exponents. These all emerge from the scaling form of the singular part
of the
free energy density

\begin{equation}\label{eq:gibbs_scal}
\mathcal{F}_s(t,h)=b^{-d}\mathcal{F}_s(b^{y_t}t,b^{y_h}h).
\end{equation}
The scaling relation (\ref{eq:gibbs_scal}) determines the properties of
$\mathcal{F}_s$ under a scale transformation of all lengths by a factor $b$.
There are two relevant parameters  in $\mathcal{F}_s(t,h)$ which control the
critical behavior: the reduced temperature $t=(T-T_c)/T_c$
and the reduced external field $h=c/c_0$. The dimensionless quantity
$\mathcal{F}_s$ is the free energy density in units of $f_\pi^2\,m_\pi^2$. All
physical information concerning the phase transition can
be encoded in the equation of state which relates the order parameter, the
external field and the reduced temperature in the vicinity of the critical
point. The leading singularities at the critical point are then completely
controlled by the thermal and the magnetic critical  exponents, $y_t$ and $y_h$.

In this section we compute critical properties and the equation of state of the
quark meson model using the FRG method and confront our results with those of
the $O(4)$ universality class. We compare the behavior of various physical
observables in the critical region with that expected from the universal scaling
function (\ref{eq:gibbs_scal}) and extract their critical exponents.

\begin{figure*}[t!]
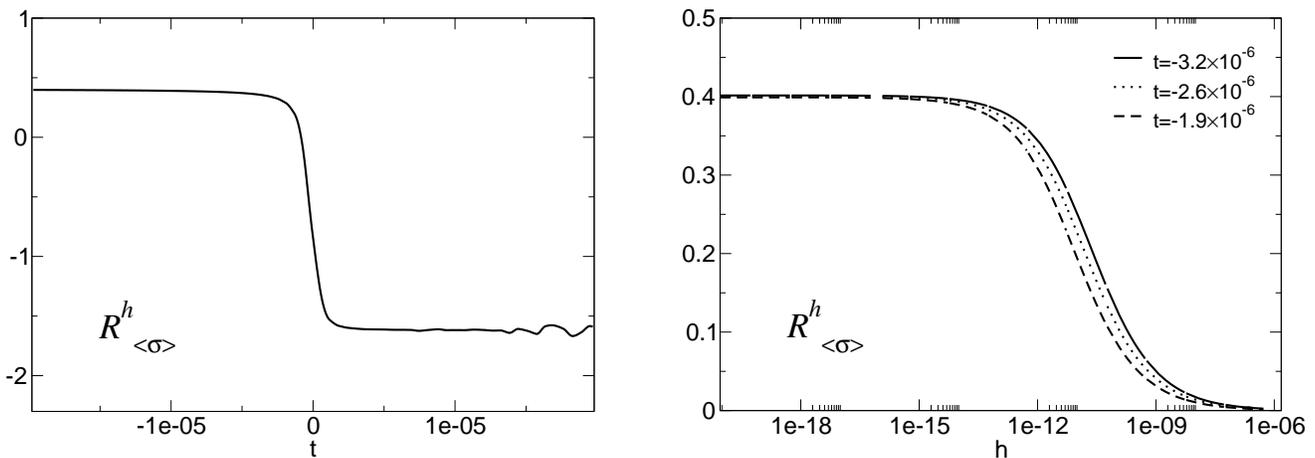

\begin{center}
\includegraphics*[height=6cm]{rbeta_t.eps}\hspace{1cm}
\includegraphics*[height=6cm]{rbeta_h.eps}
\caption{The ratio $R_{\langle\sigma\rangle}^h$  as a function of $t$ (left),
for $h=6 \times 10^{-13}$, and as a function
of $h$  for three values of the reduced temperature $t$.} \label{ratio_beta}
\end{center}
\end{figure*}

%%%%%%%%%%%%%%%%%%%%%%%%%%%%%%%
% SECTION III, order parameter
%%%%%%%%%%%%%%%%%%%%%%%%%%%%%%%
\subsection{Scaling of the order parameter}\label{subsec:crit-A}

A phase transition involving spontaneous symmetry breaking, is signalled by the
vanishing of an order parameter, when approached from the broken symmetry phase.
In the quark meson model, a possible order parameter
related to chiral symmetry breaking is obtained by differentiating the free
energy with respect to the external field $c$. Thus, it follows from Eqs.
(\ref{pot})  and (\ref{eq5}) that the order parameter can be identified with the
thermodynamic average  of the  sigma field
\begin{equation} \label{order}
\langle\sigma\rangle = \sigma_0 =\sqrt{2\rho_0}. %\sim\cond.
\end{equation}

In the FRG approach, the value of the order parameter depends on the scale $k$;
the flow of $\langle\sigma\rangle$ follows from equation Eq.~(\ref{eq:trunc2}).
The physical value of the order parameter at a given temperature $T$ is then
given by $\sqrt{2\,\rho_{k\rightarrow 0}}$, where $\rho_{k\rightarrow 0}$ is the
value obtained at the end of the flow ($k\rightarrow 0$).

In Fig.~\ref{magnet} we show the resulting dependence of the order parameter on
the reduced
temperature $t$ for several values of the external field $h$. In the chiral
limit, i.e. for  $h=0$,  the order parameter $\langle\sigma\rangle$ smoothly
decreases with increasing $t$ and finally vanishes at the critical temperature
$t=0$.  Such a behavior is characteristic of a system with a second order phase
transition. For finite $h$, the chiral symmetry of the Lagrangian is explicitly
broken and the second order phase transition turns into a crossover transition.
Thus, for small but finite $h$, the order parameter $\langle\sigma\rangle$
changes fast in a narrow temperature
interval, at the so called pseudo-critical temperature. However, in this case
the order parameter remains non-zero, even at very high temperatures, as seen in
Fig.~\ref{magnet} and there is no true phase transition.

In the vicinity of the critical point the scaling  of the order parameter
$\langle\sigma\rangle$ can be obtained from the universal form of the free
energy. Indeed, by choosing the scale factor $b$ in the Eq.
(\ref{eq:gibbs_scal}) such that
$b^{y_h}h=1$ or $b^{y_t}|t|=1$ one finds
\begin{equation} \label{defh}
{\mathcal{F}_s}=h^{d\nu_c}{\mathcal{F}_s(th^{-1/\beta\delta},1)}
\end{equation}
and
\begin{equation} \label{deft}
{\mathcal{F}_s}=|t|^{d\nu}{\mathcal{F}_s(t |t|^{-1},h|t|^{-\beta\delta})}\, ,
\end{equation}
respectively. Consequently, using
\begin{equation} \label{def}
\langle\tilde{\sigma}\rangle=-\frac{\del\mathcal{F}_s}{\del h},
\end{equation}
where $\langle\tilde{\sigma}\rangle=\langle\sigma\rangle/f_\pi$ is the reduced
order parameter,
one finds the corresponding equation of state for the magnetization
\begin{equation} \label{deffh}
\langle\tilde{\sigma}\rangle=h^{1/\delta}{F}_h(z)
\end{equation}
and
\begin{equation} \label{defft}
\langle\tilde{\sigma}\rangle=|t|^{\beta}\mathcal{F}^{'}_s(t
|t|^{-1},h|t|^{-\beta\delta}),
\end{equation}
with  $z\equiv th^{-1/\beta\delta}$ and the universal function  $F_h(z)$ is
obtained
from Eq. (\ref{defh}) by differentiation with respect to $h$. The prime on
$\mathcal{F}_s$ in (\ref{defft}) denotes the derivative with respect to the
second argument.

The well known  scaling behavior of
the magnetization
\begin{equation}\label{eq:scaling_beta_delta}
\langle\tilde{\sigma}\rangle = \left\{\begin{array}{lr} B(-t)^{\beta}, & h=0,\;
t<0 \\
B_ch^{1/\delta}, &
t=0,\; h>0 \end{array}\right.,
\end{equation}
follows from Eqs. (\ref{deffh}) and (\ref{defft}).
The constants $B_c=F_h(0)$ and $B=\mathcal{F}^{'}_s(-1,0)$ characterize the
magnetization on the so called coexistence line~\cite{Wallace} ($t<0$, $h=0$)
and at the pseudo-critical point ($t=0$, $h>0$). In the derivation of Eqs.
(\ref{defh})--(\ref{eq:scaling_beta_delta}) we have used the definitions of
the thermal and magnetic critical exponents
\begin{equation}\label{eq:def_y}
 y_t=\frac{1}{\nu},\quad y_h=\frac{1}{\nu_c}=\frac{\beta\delta}{\nu},
\end{equation}
as well as the scaling relations
\begin{equation}\label{eq:scaling}
 \delta-1=\frac{\gamma}{\beta},\quad \gamma=\nu(2-\eta),
\quad d\nu=\beta(1+\delta)
\end{equation}
and
\begin{equation}\label{eq:alpha}
\alpha=2-d\nu.
\end{equation}
In the chiral limit, the scaling behavior of the order parameter is controlled
by the critical exponent $\beta$, when the
critical temperature $t=0$ is approached from below.
Thus, $\beta$ is given by the slope in a double-logarithmic plot of $
\langle\tilde{\sigma}\rangle $ versus
$t$. The critical exponent $\delta$, on the other hand,
characterizes the scaling of  $\langle\tilde{\sigma}\rangle$ as
$h\rightarrow 0$ for $t=0$.

In Fig.~\ref{log_beta_delta} we show the scaling properties of
$\langle\tilde{\sigma}\rangle$ close
to the critical point ($t\rightarrow 0$) for vanishing  external field as well
as the scaling with $h$ at the (pseudo)critical point ($t=0$).
In both cases the order parameter
shows the expected scaling behavior. This demonstrates that, in the parameter
range considered, the free energy of the quark meson model is dominated by the
singular part Eq.~(\ref{eq:gibbs_scal}). The critical exponents extracted from a
fit to the results of Fig.~\ref{log_beta_delta} are
\begin{equation}\label{values}
 \beta\simeq 0.402  ~~~~~{\rm and}~~~~~ \delta\simeq 4.818.
\end{equation}

Both values are consistent with $O(4)$ universality. Indeed, in a
Monte Carlo simulation of the three--dimensional $O(4)$ spin
model, Kanaya and Kaya ~\cite{Kanaya:1994qe} find $\beta=0.3836(46)$ and
$\delta=4.851(22)$. The agreement with this presicion calculation is, given our
simple Ansatz, quite satisfactory. Consequently,
we conclude that the critical properties of the chiral order parameter in  the
quark--meson model are indeed governed by the $O(4)$ universality class. Our
results, obtained in the relatively simple leading order derivative expansion,
illustrate the efficiency of the
FRG approach to correctly account for a non-perturbative, long distance
physics near a phase transition.

In the mean field approximation one also finds scaling of the order parameter
$\langle\tilde{\sigma}\rangle$ close to the critical point, but with different
critical
exponents, $\beta=0.5$ and $\delta=3$. Thus, the singularity at the chiral
phase transition is substantially modified by fluctuations.

In the formulation of the flow equation for the quark-meson model we have
neglected the wavefunction renormalization for both bosonic and fermionic
fields. The good agreement with the $O(4)$ lattice results as well
as with previous FRG studies \cite{Berges:prd_qm}, where the anomalous
dimension was taken into account, indicates that the influence of the
wavefunction renormalization on the critical properties at the chiral phase
transition is quite small.

So far we have  discussed the scaling of the order parameter
$\langle\tilde{\sigma}\rangle$ with $t$ for $h=0$ in the chirally broken phase
and with $h$ at $t=0$. In the chiral limit, the order parameter vanishes
identically in the symmetric phase, i.e. for $t>0$. Thus, the order parameter
does not exhibit a singularity when the critical point is approached from above.
However, since for $h\neq 0$ the order parameter is always non-zero, it is
interesting to explore its behavior for small but finite values
of $h$, when the critical point is approached from the broken $(t<0)$ as well
as from the symmetric $(t>0)$ phase. The scaling behavior of the order parameter
is in this case characterized by a so-called effective critical
exponent \cite{Wegner}. For a given $h$, this exponent is defined by
\begin{equation}\label{lambda}
R_{\langle\sigma\rangle}^{h}:=\frac{d\log{\langle\tilde{\sigma}\rangle}}{d\log
t}=\frac{t}{\langle\tilde{\sigma}\rangle}\frac{\del
\langle\tilde{\sigma}\rangle}{\del t}.
\end{equation}
From Eqs. (\ref{eq:scaling_beta_delta}) and (\ref{lambda}) it is  clear that
for $h=0$ and $t<0$, the ratio $R_{\langle\sigma\rangle}^{h=0}$ coincides with
the critical exponent $\beta$. For $h\neq 0$, Eq.~(\ref{lambda}) can be used to
extract the leading singularity of the order parameter in the symmetric region
close to the critical point.

In the left panel of Fig.~\ref{ratio_beta} we show the resulting   dependence
of the   effective critical exponent (\ref{lambda}) on the reduced temperature
for a fixed, very small, value of the external field $h$.  The exponent
$R_{\langle\sigma\rangle}^{h}$  exhibits the expected scaling behavior. In the
broken $(t<0)$ as well as in the symmetric $(t>0)$ phase, it is essentially
independent of $t$. The effective critical exponents are different in the broken
and symmetric phases. A weak external field $h\simeq 0$ and $t<0$
corresponds to $z\to - \infty$ in Eq.~(\ref{deffh}). It follows that the
asymptotic form of the scaling
function is given by~\cite{LL} $F_h(z)\sim (-z)^\beta$ for $z\to -\infty$.
Thus, in the broken phase $(t<0)$ the exponent $R_{\langle\sigma\rangle}^{h}$
is, as
expected, 
consistent with the value of the critical exponent  $\beta $, extracted from
the scaling properties (\ref{eq:scaling_beta_delta}) of the order parameter.

In the right panel of Fig.~\ref{ratio_beta},
we show the dependence of the effective critical exponent upon $h$  in the
broken phase for three different values of $t$.
For $t<0$, the limit $h\to 0$ corresponds to $z\to - \infty$
and consequently $R_{\langle\sigma\rangle}^{h}$ again
converges to the critical exponent $\beta$. On the other hand, the limit $h\to
\infty$ corresponds to $z\to 0$. In this case,
the order parameter $\langle\tilde{\sigma}\rangle\sim
h^{1/\delta}$ is, to leading order, independent of temperature and the effective
critical exponent approaches zero. The transition between the two regimes takes
place at $|z|\sim 1$.

Also in the symmetric  phase $(t>0)$, the exponent
$R_{\langle\sigma\rangle}^{h}$ shows a power law behavior, however with a
different value of the exponent. In the transition region,
$R_{\langle\sigma\rangle}^{h}$ behaves like a smoothened step function,
interpolating between the asymptotic regions. The transition region extends from
$z\sim -1$ to $z\sim 1$. Hence the width of the step in $t$ is
expected to scale with $h^{1/\beta\delta}$. The effective critical exponent
changes from
$R_{\langle\sigma\rangle}^{h}\simeq 0.4$ in the
broken phase  to $R_{\langle\sigma\rangle}^{h}\simeq -1.5$ in the symmetric
phase. This change  in the critical behavior of the order parameter can be
understood using the universal scaling behavior of the singular part of the free
energy. To see this,
we first write the equation of state (\ref{deffh}) and (\ref{defft}) in the
Widom--Griffiths form
\begin{equation}\label{widom1}
y=f(x),
\end{equation}
where
\begin{equation}\label{widom2}
 {y}\equiv {\frac{h}{\langle\tilde{\sigma}\rangle^\delta}} ~~ {, }~~
x\equiv {\frac{t}{\langle\tilde{\sigma}\rangle^{1/\beta}}}.
\end{equation}

The scaling relations (\ref{eq:scaling_beta_delta}) are, modulo constant
factors, obtained if $f(0)=1$ and $f(-1)=0$. The constant factors are recovered
by a simple rescaling of $t$ and $h$. A comparison of the Widom-Griffiths form
of
the equation of state (\ref{widom1}) with that
given in Eq. (\ref{deffh}), shows that the following relations hold:
\begin{equation}\label{relation}
F_h(z)=f(x)^{-1/\delta} ~~,~~z=xy^{-1/\beta\delta}.
\end{equation}
Consequently, Eq. (\ref{deffh}) can be rewritten as
\begin{equation} \label{wdeffh}
\langle\tilde{\sigma}\rangle=B_c h^{1/\delta}f(x)^{-1/\delta},
\end{equation}
where we have reinstated the constant factor $B_c$.
This equation of state can be used to explore the scaling properties of
$\langle\tilde{\sigma}\rangle$ when approaching the critical point from the
symmetric phase. For $t>0$ and for $h\to 0$, the chiral order parameter
(normalized to its value in vacuum) is very small. Consequently, the argument
$x$ of the function $f(x)$ in equation (\ref{wdeffh}) is  very large and the
scaling behavior of $\langle\tilde{\sigma}\rangle$ for $t>0$ and for small but
finite $h$ is determined by the  asymptotic form of the scaling function $f(x)$
for $x\to \infty$. In this
limit, $f(x)$  is given by Griffiths' analyticity
condition ~\cite{Griffiths},
\begin{equation}\label{eq:large_x}
f(x)=\sum_{n=1}^{\infty}f_n x^{\gamma-2(n-1)\beta}.
\end{equation}
Retaining only the leading term, $ f(x)\simeq x^\gamma$,
one finds that for $t>0$ and $h\to 0$ the equation
of state scales as
\begin{equation}\label{eq:scal_cond_t}
\langle\tilde{\sigma}\rangle \sim t^{-\gamma}h.
\end{equation}
Thus, for $t>0$ and $h\neq 0$, we expect $R_{\langle\sigma\rangle}^{h} =
-\gamma$, in good agreement with the results shown in Fig.~\ref{ratio_beta}.
This result implies~\cite{LL} that for $z\to \infty$, $F_h(z) \simeq
z^{-\gamma}$. We note that in the chiral limit ($h=0$), the order parameter
vanishes identically for $t>0$, as expected.

From the scaling
behavior of $\langle\tilde{\sigma}\rangle$ shown in the left panel of
Fig.~\ref{ratio_beta}, we find $\gamma\simeq 1.6$. This
is consistent with the lattice result for the $O(4)$ spin system obtained in
ref.~\cite{Kanaya:1994qe}, $\gamma\simeq 1.5$. Thus, the
FRG method applied to the quark meson model yields a scaling behavior of the
order parameter, which is consistent with that found for the $O(4)$
universality class. This is the case at the critical point, on the coexistence
line and when approaching the critical point from the high temperature phase for
a finite value of the external magnetic field.

\begin{figure}[t!]
\begin{center}
\includegraphics*[width=8.6cm]{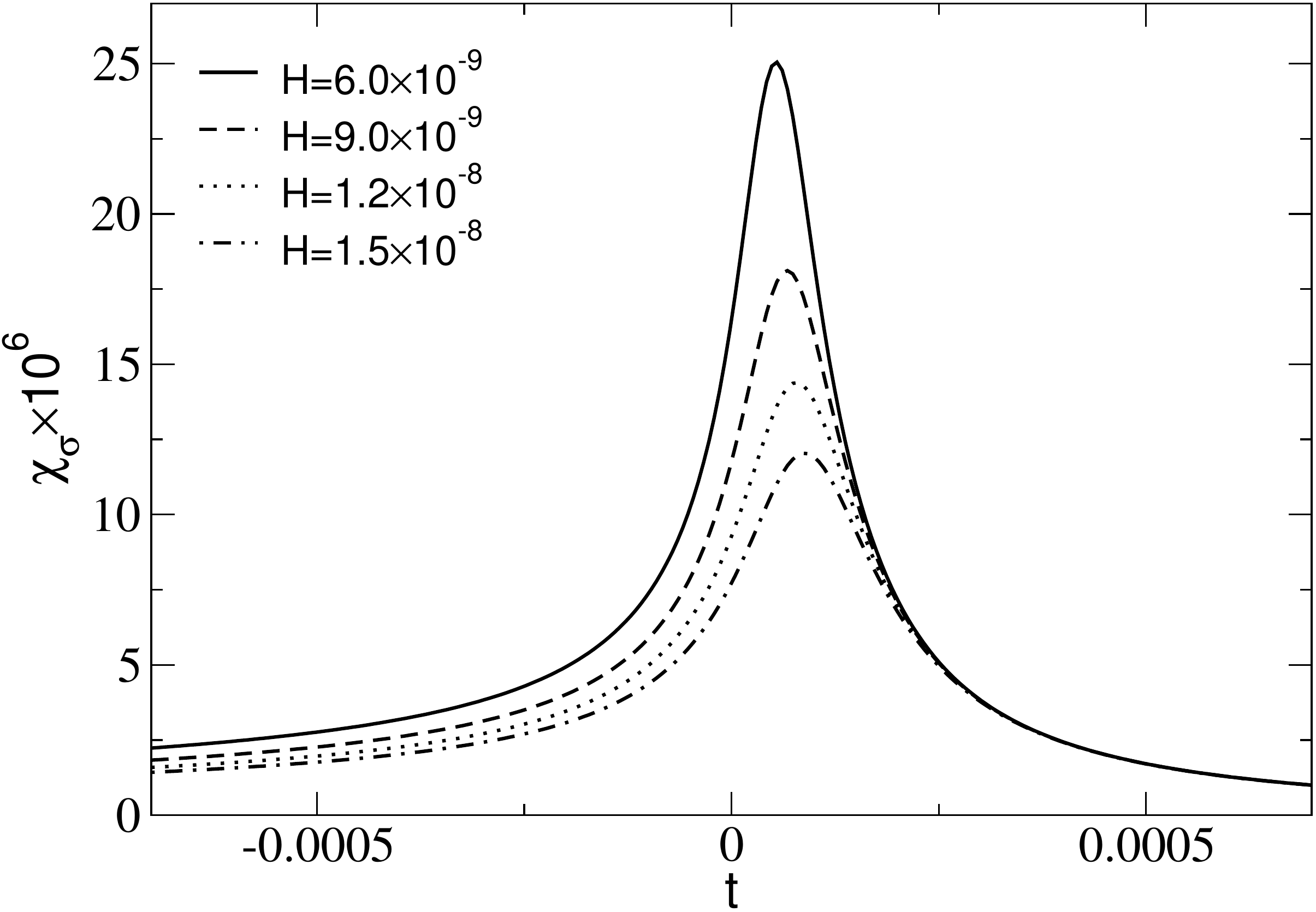}
\caption{The susceptibility $\chi_{\sigma}$ as a function of the reduced
temperature $t$
for several values of the field $h$.} \label{sus_peak}
\end{center}
\end{figure}

\begin{figure*}
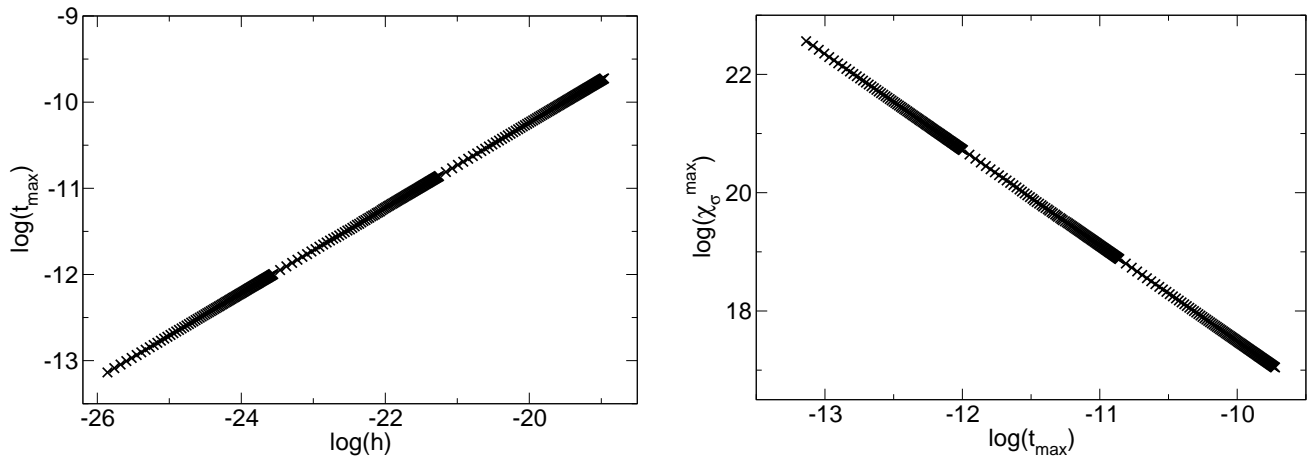

\begin{center}
\includegraphics*[height=6cm]{t_h_max.eps}\hspace{5mm}
\includegraphics*[height=5.9cm]{sus_h_max.eps}
\caption{Scaling of the pseudo-critical temperature defined by the maximum of
the susceptibility $\chi_{\sigma}$ (left) and of the maximum value of
$\chi_{\sigma}$ (right).} \label{sus_t_scal}
\end{center}
\end{figure*}

%%%%%%%%%%%%%%%%%%%%%%%%%%%%%%%
% SECTION III, part susceptibility
%%%%%%%%%%%%%%%%%%%%%%%%%%%%%%%
\subsection{Scaling of the chiral susceptibilities}

The critical properties of a thermodynamic system can be explored by studying
the fluctuations of various observables. In particular, the fluctuations of the
order parameter probe  the order of the phase transition and the
position of a possible critical end point. In statistical physics, fluctuations
are reflected in the corresponding susceptibilities $\chi$. In the case of the
chiral order parameter, the corresponding susceptibility is defined as the
response to a change of the external field $h$; the susceptibility $\chi$ is
obtained by taking the second derivative of the effective potential with respect
to $h$. The susceptibility in other channels is computed analogously. It follows
that the susceptibilities are inversely proportional to the corresponding mass
squared. Consequently, the divergence of a susceptibility, e.g. at the critical
end point, signals a zero-mass mode of the corresponding effective field.

In the chiral quark-meson model, we are not only dealing with fluctuations of
the order parameter, i.e. of the sigma field, but also with a pseudo-Goldstone
mode, corresponding to fluctuations of the pion field. The sigma and the pion
represent the longitudinal and transverse modes of the $O(4)$ field $\phi$,
respectively. One therefore distinguishes between the longitudinal ($\sigma$)
and transverse ($\pi$) order parameter susceptibilities.

The longitudinal susceptibility is given by the derivative of the
order parameter with respect to the external field
\begin{equation}\label{eq:sus_sigma}
\chi_l=\chi_{\sigma}=\frac{\partial\langle\tilde{\sigma}\rangle}{\partial h},
\end{equation}
whereas  the transverse susceptibility, which is directly related to the
fluctuation of
the Goldstone modes, is obtained from
\begin{equation}\label{eq:sus_pi}
\chi_t=\chi_{\pi}=\frac{\langle\tilde{\sigma}\rangle}{h}.
\end{equation}
The latter expression is a direct consequence of the $O(4)$ invariance of the
free energy for $h=0$ and follows from the corresponding Ward
identity~\cite{bre}.

\begin{figure}[b!]
\begin{center}
\includegraphics*[width=8.6cm]{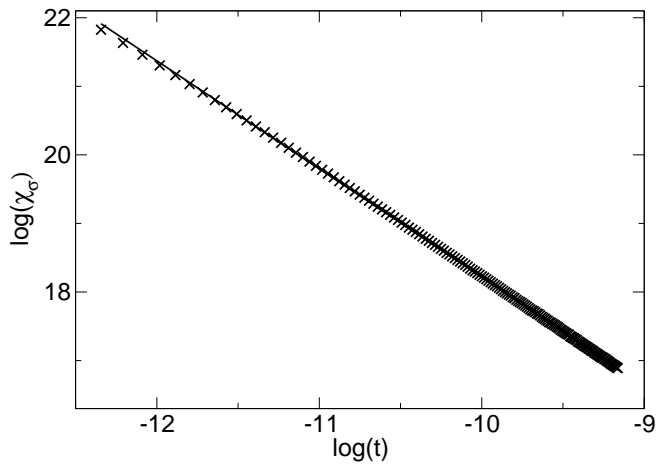}
\caption{Double logarithmic plot of the longitudinal susceptibility in the
symmetric phase for a small $h$ as a function the reduced temperature.}
\label{log_gamma}
\end{center}
\end{figure}

The critical  behavior of the longitudinal and transverse chiral susceptibility
can be obtained from the scaling form of the singular  part of the free energy
or directly from the magnetic equation of state (\ref{wdeffh}).

\begin{figure*}
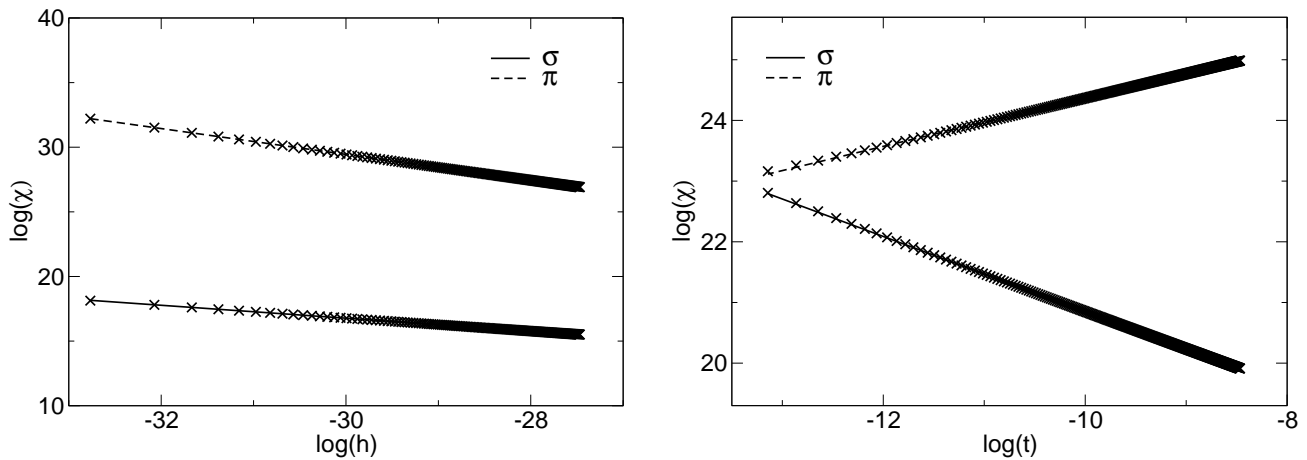

\begin{center}
\includegraphics*[height=6cm]{log_sp_h.eps}\hspace{5mm}
\includegraphics*[height=5.9cm]{log_sp_t.eps}
\caption{Scaling of $\chi_{\sigma}$ and $\chi_{\pi}$ in the broken phase
($t<0$) as functions of $h$ (left) and $t$ (right)} \label{log_gamma_th}
\end{center}
\end{figure*}

In Fig.~\ref{sus_peak} we show the longitudinal susceptibility $\chi_{\sigma}$
as a function of the reduced temperature for several values of the symmetry
breaking field $h$. At finite $h$, this susceptibility shows a peak structure
with a maximum at the pseudo-critical temperature $t_{\mathrm{max}}(h)$. With
decreasing  $h$, there is a shift of $t_{\mathrm{max}}$ and an increase of the
fluctuations at the pseudo-critical point; in the chiral limit $\chi_{\sigma}$
diverges at the transition temperature, because the mass of the sigma field
vanishes.

Using the results shown in the Fig.~\ref{sus_peak}, one can define a
pseudo-critical line in the ($t,h$)--plane that characterizes the dependence of
the transition temperature $t_{\mathrm{max}}$ on the symmetry breaking field
$h$. Renormalization group arguments imply that the pseudo-critical temperature
and the maximum of the longitudinal susceptibility should exhibit universal
scaling according to
~\cite{Pelissetto:2000ek}:
\begin{equation}\label{eq:sus_t_scal}
t_{\mathrm{max}}(h)\simeq T_ph^{1/(\gamma+\beta)},\quad
\chi_{\sigma}(t_{\mathrm{max}},h)\simeq C_pt_{\mathrm{max}}^{-\gamma},
\end{equation}
with $O(4)$ critical exponents.

In Fig.~\ref{sus_t_scal} we confront the scaling behavior of the susceptibility
with Eq.~(\ref{eq:sus_t_scal}). The critical exponents are extracted by
performing a linear logarithmic fit to the pseudo-critical temperature and the
maximum value of the susceptibility. We find $1/(\gamma+\beta)=0.494$ and
$\gamma=1.618$, which implies $\beta=0.406$, in good agreement with the results
obtained above using Eqs.~ (\ref{eq:scaling_beta_delta}) and
(\ref{eq:scal_cond_t}). Thus, the scaling of
of $\chi_{\sigma}$ with the symmetry breaking field $h$ is also consistent with
$O(4)$ universality.

In the symmetric phase, the longitudinal and the transversal susceptibilities
coincide $(\chi_{\sigma}=\chi_{\pi})$. Consequently, the strength of the
singularities at the critical point $t=0$ are controlled by the same
critical exponents.

To find  the scaling behavior of $\chi$ when approaching the critical point
from the symmetric phase we use the Widom--Griffiths form of the magnetic
equation of state (\ref{wdeffh}) together with the large-$x$ expansion of the
scaling function
$f(x)$.

From Eqs. (\ref{eq:scal_cond_t}), (\ref{eq:sus_sigma}) and (\ref{eq:sus_pi})
it is clear that for $h\to 0$ in the symmetric phase
($t>0$)
\begin{equation}\label{eq:scal_sus}
 \chi_\sigma=\chi_\pi\simeq t^{-\gamma}.
\end{equation}
Thus, when approaching the critical point from the symmetric phase, the scaling
behavior of the longitudinal and transverse susceptibility is controlled by the
critical exponent $\gamma$.

In Fig.~\ref{log_gamma} we show the scaling behavior of the susceptibility for
$t>0$ and in the limit of $h\to 0$. From the slope of the line we obtain
$\gamma=1.575$, in good agreement the value obtained from the analysis of the
scaling of the chiral susceptibility $\chi_{\sigma}$ following
Eq.~(\ref{eq:sus_t_scal}). The small difference is may be due to the
uncertainty in the determination of the critical temperature.

In the broken phase, on the coexistence line, the critical behavior of the
longitudinal and transverse susceptibilities differ. In the chiral limit, the
order parameter is finite  in the broken phase as long as $t<0$. Thus, for
$T<T_c$ the transverse susceptibility (\ref{eq:sus_pi}) diverges in the chiral
limit, $\chi_{\pi}\sim h^{-1}$, due to the appearance of Goldstone bosons.
A less obvious result is the divergence of the longitudinal susceptibility on
the coexistence line, where to leading order $\chi_{\sigma}\sim
h^{-1/2}$~\cite{Wallace,Engels:2003nq}.

The scaling properties of the susceptibilities
can be obtained directly from the Widom-Griffiths form of the magnetic equation
of state. Indeed, using Eqs. (\ref{wdeffh}), (\ref{eq:sus_sigma}) and
(\ref{eq:sus_pi}) one finds

\begin{equation}\label{eq:scal_sus_sigma}
\chi_{\sigma}=B_c h^{1/\delta-1} \frac{\beta f(x)^{1-1/\delta}}{\beta\delta
f(x)-xf'(x)},
\end{equation}
for the longitudinal and
\begin{equation}\label{eq:scal_sus_pi}
\chi_{\pi}=B_c h^{1/\delta-1}f(x)^{-1/\delta},
\end{equation}
for the transverse susceptibility.

In the vicinity of the coexistence line, where $x\to -1$, the scaling function
is, in a three-dimensional system, found to be of the
form~\cite{Wallace,brezin-wallace}
\begin{equation}\label{eq:x_1}
f(x)\approx c_f(1+x)^2\,.
\end{equation}
Using (\ref{eq:x_1}) in (\ref{eq:scal_sus_sigma}) and (\ref{eq:scal_sus_pi}),
one then finds the critical behavior of the 
susceptibilities for $t<0$ and $h\to 0$:
\begin{equation}\label{eq:scal_sigma}
\chi_{\sigma}\sim h^{-1/2}(-t)^{\beta-(\beta\delta)/2}
\end{equation}
and
\begin{equation}\label{eq:scal_pi}
\chi_{\pi}\sim h^{-1}(-t)^{\beta}\, .
\end{equation}
The divergence of the susceptibilities on the coexistence line is due to the
transverse modes~\cite{Wallace}, whose mass term is proportional to the
symmetry-breaking term, $m_\pi^2 \sim h$. The different exponents for
$\chi_{\sigma}$ and $\chi_{\pi}$ reflect the direct and indirect dependence of
the susceptibilities on the soft transverse modes. The critical behavior of
the longitudinal susceptibility close to the coexistence line, fixes the
exponent of the
subleading term of the scaling function defined in (\ref{deffh}) for $z<0$
\begin{equation}
F_h(z)\sim (-z)^\beta (1+C (-z)^{-\beta\delta/2})\, ,
\end{equation}
where $C$ is a constant.

In Fig.~\ref{log_gamma_th} we show log--log plots of the longitudinal and
transverse susceptibilities, obtained by solving the flow equations for
the quark meson  model (\ref{eq:rgflow}), as functions of $h$  and $t$ near the
coexistence line. A linear fit to the results shown on the left
yields $0.499$ and $1.000$ for the critical exponents of $\chi_\sigma$ and
$\chi_\pi$, in excellent agreement with the expected scaling
behavior with $h$ implied by Eqs. (\ref{eq:scal_sigma}) and (\ref{eq:scal_pi}).
The corresponding fit to the $t$-dependence, shown in the right plot, gives
$(\beta-\beta\delta/2)=-0.617$ and $\beta=0.395$, which implies $\delta =
5.124$, in fair agreement with our previous results for the critical exponents
$\beta$ and $\delta$ obtained from the scaling of the order parameter.

In order to exhibit the scaling behavior of the susceptibilities
in a broad  parameter range, which covers both the broken and symmetric phases,
we introduce the ratios
\begin{equation}\label{eq:ratio_gamma}
R_{\chi_{\pi,\sigma}}^h=\frac{t}{\chi_{\pi,\sigma}}
\frac{\del\chi_{\pi,\sigma}}{\del t}
\end{equation}
for the transverse $\chi_\pi$ and the longitudinal  $\chi_\sigma$
susceptibilities. The ratios $R_{\chi_{\pi,\sigma}}^h$ define effective
critical exponents, analogous to Eq.~(\ref{lambda}).

\begin{figure*}[t]
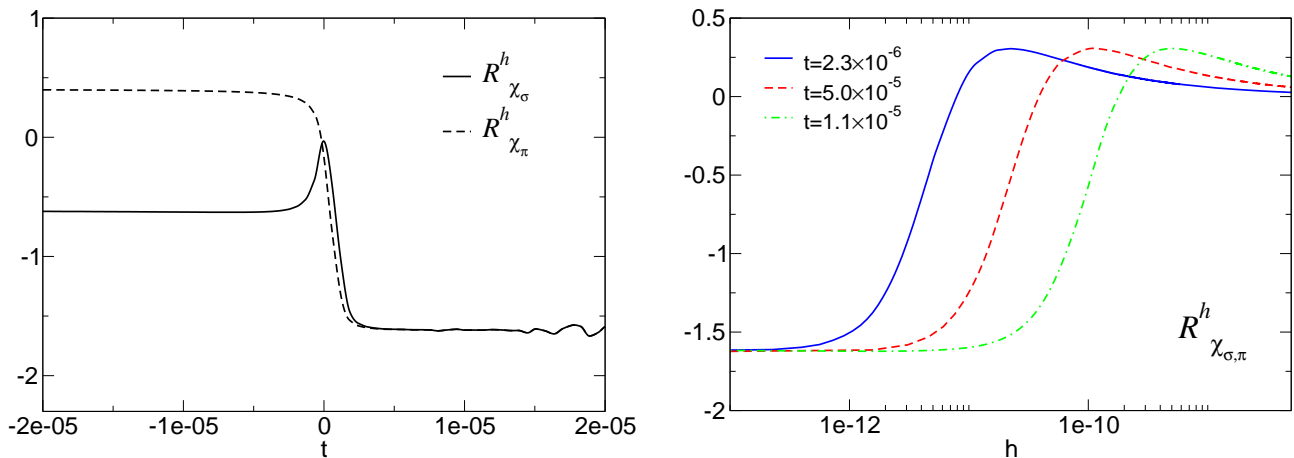

\begin{center}
\includegraphics*[height=6cm]{rgamma_t.eps}\hspace{5mm}
\includegraphics*[height=6cm]{rgamma_h.eps}
\caption{The ratios $R_{\chi_{\pi,\sigma}}^t$ as functions of $t$ for a
small $h$ (left panel) and as functions of $h$ at three values of the
temperature in the symmetric phase ($t>0$) (right  panel).} \label{ratio_gamma}
\end{center}
\end{figure*}

In  Fig.~\ref{ratio_gamma} we summarize the $t$ and $h$ dependence of the
$R_{\chi_\pi}^h$ and $R_{\chi_\sigma}^h$ in the vicinity to the critical point.
On the left we show how the critical exponents of the susceptibilities
interpolate between the broken and symmetric phases, for fixed  $h$, while on
the right the $h$--dependence of $R_{\chi_{\pi,\sigma}}^h$ in the symmetric
phase is illustrated for a few values of $t$.

In the symmetric phase, $\chi_\pi$ and $\chi_\sigma$ coincide and their
critical behavior is determined by the same exponent $(\gamma\sim 1.58)$, as
seen in Fig.~\ref{ratio_gamma}--left. This corresponds to the limit $z\to
\infty$, and  is consistent with the effective critical exponent obtained for
the order parameter for $t>0$. The dependence upon $h$ in the symmetric phase,
shown on the right, is also identical for the longitudinal and transverse
susceptibilities. Here the $z\to \infty$ limit corresponds to $h\to 0$ for fixed
$t>0$. Thus, also in this limit, $R_{\chi_{\pi,\sigma}}^h\to -\gamma$, as shown
in Fig.~\ref{ratio_gamma}--right. Again, the $h\to \infty$ limit corresponds to
$z\to 0$, where the order parameter and consequently also the susceptibility
approaches a temperature independent function of $h$ and the corresponding
effective critical exponent converges to zero with increasing $h$. The
transition between the two regimes again takes place at $z\simeq 1$.

In the broken phase (see Fig.~\ref{ratio_gamma}--left), the critical exponents
differ, $R_{\chi_{\sigma}}^h\simeq\beta(1-\delta/2)$ and $R_{\chi_{\pi}}^h\simeq
\beta$, in complete agreement with Eqs.~(\ref{eq:scal_sigma}) and
(\ref{eq:scal_pi}).

In section \ref{subsec:crit-A} we determined the critical exponent $\delta$ by
studying the scaling of the order parameter $\langle\tilde{
\sigma}\rangle$ with $h$ for $t=0$. We note that this critical exponent may be
determined in an alternative way, by exploiting the scaling properties of the
susceptibilities~\cite{Boyd, Karsch:1994hm, Kocic:1992is} $\chi_\pi$ and
$\chi_\sigma$. Consider the ratio
\begin{equation}\label{eq:delta}
\Delta(t,h)=\frac{\chi_{\pi}^{-1}}{\chi_{\sigma}^{-1}}=
\frac{h}{\langle\tilde{\sigma}\rangle}\frac{\partial\langle\tilde{\sigma}
\rangle}{\partial h}\, ,
\end{equation}
which has the useful property that its value at $t=0$ is independent of $h$
within the scaling region. Indeed, using the parametrization of the
susceptibilities (\ref{eq:scal_sus_sigma}) and (\ref{eq:scal_sus_pi}) as well as
the properties of the Widom-Griffiths scaling function $f(x)$ discussed above,
one finds
\begin{equation}
\lim_{h\to 0}\Delta(t,h)=\left\{
\begin{array}{lr} 1, & t>0\, , \\
                            1/\delta, & t=0 \, ,\\
                            0, & t<0\, .\\
\end{array}\right.
\end{equation}

Consequently, the ratio $\Delta(t,h)$ can be used to directly extract the value
of the critical exponent $\delta$, without resorting to logarithmic fits.
Moreover, this ratio simultaneously reveals the position of the
critical point, since in the scaling regime, all $\Delta(t,h)$, plotted as
functions of $t$ for different $h$, cross at $t=0$.

\begin{figure*}
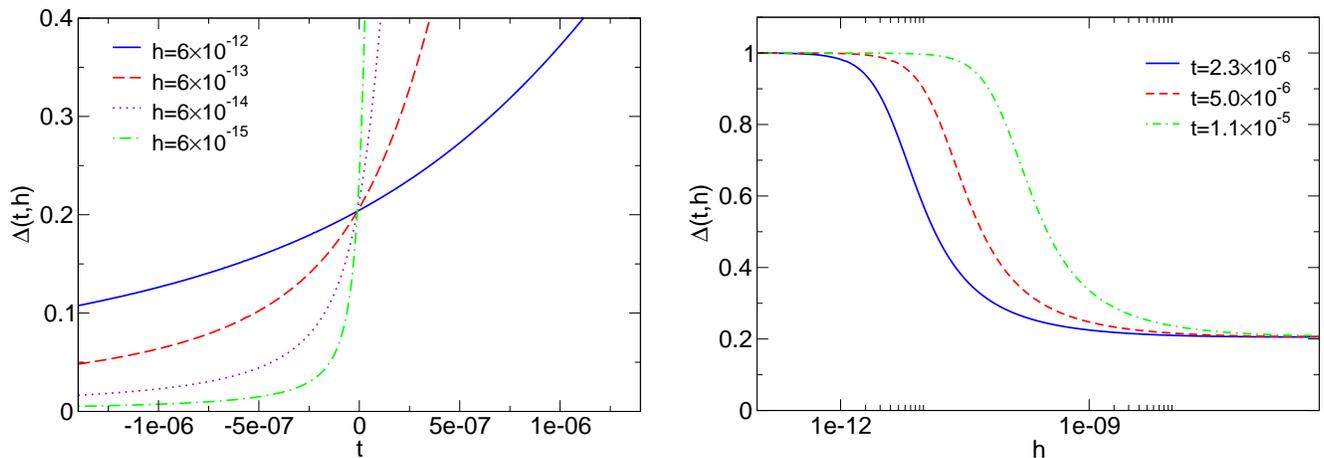

\begin{center}
\includegraphics*[height=6cm]{cross_d.eps}\hspace{5mm}
\includegraphics*[height=5.9cm]{d_h.eps}
\caption{The ratio $\Delta(t,h)$ defined in  Eq.~(\ref{eq:delta}) as a function
of the reduced temperature  $t$ and the external field $h$. In the left--hand
figure the curves cross in the point $(0, 1/\delta)$.} \label{delta_h}
\end{center}
\end{figure*}

In Fig.~\ref{delta_h}--left we show $\Delta(t,h)$ as a function of $t$ for four
values of $h$. As expected all lines cross at $t=0$, yielding the value of the
critical exponent $\delta\approx 4.91$. This value is in good agreement with the
one obtained above, using a logarithmic fit to the dependence of the order
parameter on $h$.

In the right panel of Fig.~\ref{delta_h} we show the scaling of $\Delta$ as a
function of $h$, for fixed $t>0$. In the chiral limit, that is for $h= 0$, the
longitudinal and transverse susceptibilities coincide above the critical
temperature. Consequently  $\Delta=1$ in the limit $h\to 0$, as shown in the
plot. In the opposite limit, i.e. for $h\to \infty$, we find $\Delta =
1/\delta$. The coincidence of the two limits $h\to\infty$ and $t\to 0$ follows
from the explicit form for this ratio in terms of the scaling equation of state
(\ref{deffh}), $\Delta=1/\delta - (z/\beta\delta)(F_h^\prime(z)/F_h(z))$. Since
the limits $t\to 0$ for finite $h$ and $h\to \infty$ for finite $t$ both imply
$z=t h^{-1/\beta\delta}\to 0$, the result $\Delta=1/\delta$ is obtained in both
cases.

%%%%%%%%%%%%%%%%%%%%%%%%%%%%%%%%%%%%%%%%%%
%SECTION III, correlation length
%%%%%%%%%%%%%%%%%%%%%%%%%%%%%%%%%%%%%%%%%%
\subsection{ Correlation lengths and their critical behavior}

A correlation length $\xi$ governs the exponential decay of the corresponding
correlation function with increasing separation. The critical behavior of
correlation lengths is an important characteristic of phase transitions. At the
critical point of a second-order phase transition, the order-parameter
correlation length diverges, heralding the onset of a long-range order.

The decay of a correlation function is controlled by the relevant mass
scale in the system. In a chiral model, such as the one considered here, there
are two such scales, the transverse mass connected with the pion and the
longitudinal mass related to the sigma field. Consequently, like in the case of
magnetization, one introduces a transverse and a longitudinal
correlation length $\xi_\pi$ and $\xi_\sigma$. In the critical region for
$t>0$, the
correlation lengths are directly related to the corresponding
susceptibilities \cite{Engels:2003nq,Fisher}
\begin{equation}\label{relation-1}
\chi_l\sim\xi_l^2~~,~~\chi_t\sim\xi_t^2.
\end{equation}
Consequently, in the chiral quark-meson model, the correlation lengths are
determined by the pion and sigma masses
\begin{equation}
\xi_{\sigma}\sim\frac{1}{m_{\sigma}}, \quad
\xi_{\pi}\sim\frac{1}{m_{\pi}}.
\end{equation}

In the broken symmetry phase, the soft transverse modes,
which renormalize the longitudinal susceptibility in a non-trivial way, as
shown in Eq.~\ref{eq:scal_sigma}, also profoundly influence the longitudinal
correlation length~\cite{fisher-barber-jasnow}. This modifies the relation
between the correlation length and the susceptibility in the longitudinal
channel for $t<0$. We do not address this problem here.

\begin{figure}[t!]
\begin{center}
\includegraphics*[width=8.6cm]{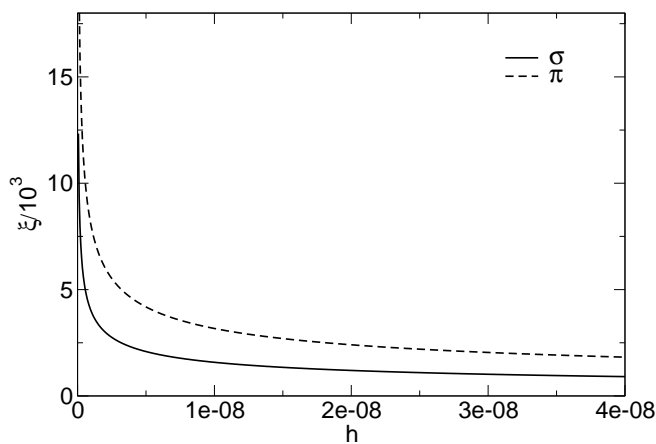}
\caption{The correlation lengths $\xi_{\sigma}$ and $\xi_{\pi}$ at the critical
point $t=0$ as functions of the external field $h$.}
\label{corr_length}
\end{center}
\end{figure}

In Fig.~\ref{corr_length} we show the correlation lengths,  $\xi_{\sigma}$ and
$\xi_{\pi}$, at the critical point $t=0$ as functions of the external field $h$.
As expected both the longitudinal and transverse correlation lengths
diverge as $h\to 0$. The divergence of $\xi_\sigma$ reflects the softening of
order parameter fluctuations as the critical point is approached, while that of
$\xi_\pi$ signals the appearance of Goldstone bosons at the second order phase
transition.

The scaling of the correlation length in the critical region is  obtained
from the
scaling function (\ref{deffh})  and the relation (\ref{relation-1}).
On the critical line $(t=0, h\to 0)$, the transverse
and longitudinal correlation length exhibit the same
singularity, controlled by the exponent
$\nu_c$ ~\cite{Engels:2003nq}
\begin{equation}
\xi_{\sigma,\pi}\sim h^{-\nu_c}\, .
\end{equation}
Their ratio, $\xi_\pi/\xi_\sigma$, is at
$t=0$ approximately
equal to 2 (see Fig.~\ref{corr_length}). Thus, the value expected for this ratio
in the broken symmetry
phase~\cite{fisher-barber-jasnow} is obtained also on the critical
line, in agreement with~\cite{Engels:2003nq}.

\begin{figure*}
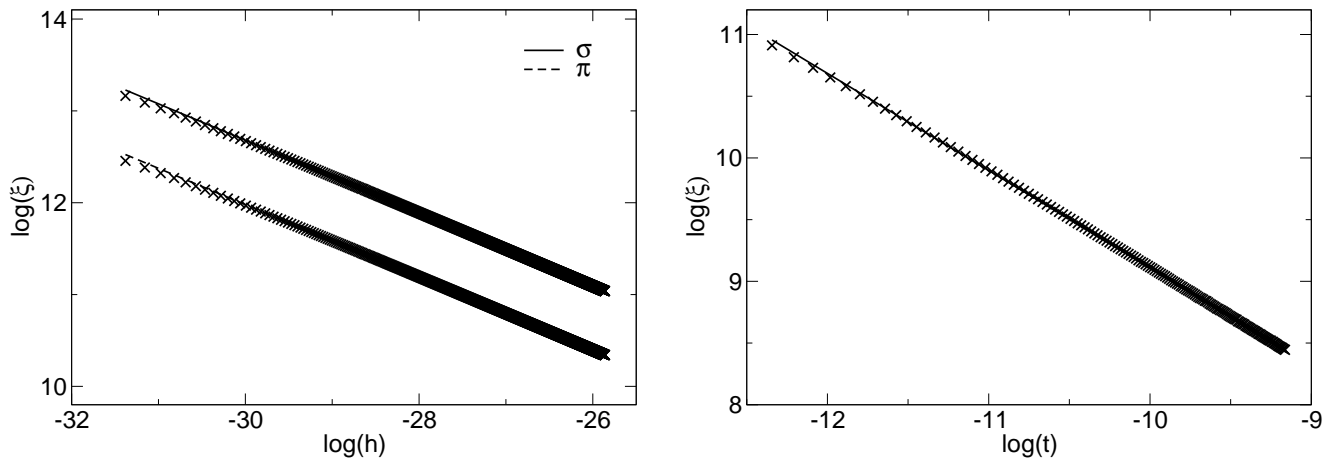

\begin{center}
\includegraphics*[height=6cm]{log_nuc.eps}\hspace{5mm}
\includegraphics*[height=6cm]{log_nu.eps}
\caption{The correlation lengths $\xi_{\sigma}$ and
$\xi_{\pi}$ as functions of $h$ at $t=0$ (left) and as functions of $t$ at
$h=0$ (right). As discussed in the text, the scaling of the correlation lengths
is governed by the critical exponents $\nu_c$ and $\nu$. Note that the
normalization of the correlation lengths
differ from that used in Fig.~\ref{corr_length}.}
\label{log_nuc_nu}
\end{center}
\end{figure*}

In the symmetric phase ($t>0$), the correlation lengths $\xi_{\sigma}$ and
$\xi_{\pi}$ coincide and the scaling with $t$ is governed by the critical
exponent $\nu$
\begin{equation}
\xi_{\sigma,\pi}\sim t^{-\nu}.
\end{equation}
In  Fig.~\ref{log_nuc_nu} we show the scaling behavior of the correlation
lengths on the
critical line and in the symmetric phase. A fit of the critical exponents
yields  $\nu_c=0.396$  and $\nu=0.787 $, respectively.

%%%%%%%%%%%% critical exponents summary %%%%%%%%%%%%%%%%

Our results for the various critical exponents are summarized in
Table~\ref{tab:exponents}. We give the values obtained in the logarithmic
fits shown in Figs.~\ref{log_beta_delta},\ref{log_gamma} and
\ref{log_nuc_nu}, which we deem to be more accurate than the other
determinations. The value for the coefficient $\alpha$ is obtained by using
the hyperscaling relation (\ref{eq:alpha}). Our results are
compared with recent lattice results and mean-field theory exponents for the
$O(4)$ model. The mean-field values clearly deviate from
the FRG and lattice results, showing the importance of fluctuations near the
critical point. Our exponents, obtained with the FRG approach,
are in good agreement with the Monte Carlo lattice results~\cite{Kanaya:1994qe}
in spite of the fact, that we have neglected the anomalous dimension in
the flow equations.

The finding that the anomalous dimension is fairly
unimportant for the critical behavior, is consistent with the relatively small
value of the corresponding critical exponent $\eta$ in the $O(4)$ universality
class. In spite of the fact that we neglect momentum dependent couplings
in the RG flow, a fairly reasonable value for the critical exponent $\eta$, can
be extracted by means of the hyperscaling relation
\begin{equation}
 \delta=\frac{d+2-\eta}{d-2+\eta}.
\end{equation}
Using our value for $\delta$, we find $\eta\approx0.031$, which should be
compared with the lattice result, $\eta=0.0254$~\cite{Kanaya:1994qe}. We note,
however, that using the middle scaling relation in (\ref{eq:scaling}) to
determine $\eta$ fails; the resulting value is small and negative. This
illustrates the uncertainty introduced by the leading order derivative
expansion.

%\begin{table}
%\begin{ruledtabular}
%\begin{tabular}{ccccc}
%& $\beta_{\mathrm{eff}}$ & $\gamma_{\mathrm{eff}}$ &$\nu_{\mathrm{eff}}$ &
%$(\beta-\beta\delta/2)_{\mathrm{eff}}$  \\
%\hline
%broken phase ($t<0$)& 0.395 & 1.612 & 0.806 & \\
%symmetric phase ($t>0$) & 0.3836 & 1.477 & 0.7479 & \\
%\end{tabular}
%\end{ruledtabular}
%\caption{Effective critical exponents and its numerical values}
%\end{table}

\begin{table*}
\caption{The critical exponents of the $O(4)$ model obtained in lattice
calculations, in the mean-field approximation and in the present FRG approach.}
\begin{ruledtabular}
\begin{tabular}{lcccccc}
% &\multicolumn{2}{c}{$D_{4h}^1$}&\multicolumn{2}{c}{$D_{4h}^5$}\\
 & $\beta$ & $\delta$ & $\gamma$ & $\nu$ & $\nu_c$ & $\alpha$  \\ \hline
 $O(4)$ lattice~\cite{Kanaya:1994qe}
 & 0.3836 & 4.851 & 1.477 & 0.7479 & 0.4019\footnote{Obtained using
the definition of $\nu_c$ (\ref{eq:def_y}).} & -0.244\\
  Mean Field & 0.5 & 3 & 1 & 0.5 & 1/3 & 0 \\
 FRG (our work) & 0.402 & 4.818 & 1.575 & 0.787 & 0.396 &
-0.361$\footnote{Obtained using the scaling relation (\ref{eq:alpha})}$
\\
\end{tabular}\label{tab:exponents}
\end{ruledtabular}
\end{table*}

%%%%%%%%%%%%%%%%%%%%%%%%%%%%%%%%%%%%%%%%%%%%%%%%%%%%%%%%%%%%%%%
\section{Summary and Conclusions}\label{sec:concl}

We have discussed the critical properties of the chiral quark-meson model
at the finite-temperature phase transition. The
focus was on the universal scaling  behavior of various physical quantities.
Fluctuations and non-perturbative
effects were accounted for by employing the Functional
Renormalization Group (FRG) method. We derived the flow equation for the scale
dependent thermodynamic potential at finite temperature and chemical potential
in the presence of an external field. The flow equation was then solved
numerically, by employing a truncated polynomial expansion around the minimum of
the effective action.
We have shown that the choice of regulator functions (\ref{eq:BRb2}) and
(\ref{eq:BRf2}) leads to transparent expressions and reliable numerical results.
In this work, we have neglected the bosonic and fermionic wave function
renormalization and the running of the Yukawa coupling. We argue that, in the
critical region, these effects lead to sub-leading corrections to the scaling
behavior.

We have computed various thermodynamical quantities that exhibit critical
behavior near
the chiral phase transition. In particular, we have analyzed the scaling
behavior of the chiral order parameter, its longitudinal and transverse
susceptibilities as well as the correlation lengths in the chiral limit and in
the presence of an  external symmetry breaking field.
We found that the chiral order parameter and its susceptibilities scale
near the chiral phase transition following  the universal scaling behavior of
the singular part of the free energy. Here we have extracted the corresponding
critical exponents at the critical point and at the coexistence line and have
also analyzed the scaling of the pseudocritical temperature and the maxima of
the longitudinal susceptibility in an external field.

We have also analyzed the effective critical exponents, which govern the leading
scaling behavior above and below the critical temperature for a system in an
external symmetry breaking field. In this context, the longitudinal
and transverse susceptibilities are of particular interest, since they diverge
at the coexistence line. The effective critical exponents obtained numerically,
are in full agreement with the Widom-Griffiths form of the magnetic equation of
state.

The critical exponents obtained in this work within the quark meson model
by employing the FRG method, are in  a very good agreement with the
recent lattice results for the $O(4)$ spin system. This confirms the
expectation that the chiral phase transition of the 2--flavor effective chiral
Lagrangian belongs to the $O(4)$ universality class. Furthermore, our
results demonstrate the power of the FRG approach  and the validity of the truncation
 approximation to the flow equation. This approach offers an efficient
framework for describing fluctuations and non-perturbative long-distance
dynamics near the chiral phase transition.

We note in closing, that the critical region (the region of
true non-trivial critical behavior) of the chiral quark-meson model turns out to
be very small in our calculation. This is reflected in the very small values for
$t$ and $h$ needed in order to obtain stable results in the logarithmic fits of
the critical exponents. Our finding is consistent with the suppression of the
Ginsburg region~\cite{kogut-stephanov-strouthos} in $O(N)$ models for large $N$.
However, because the size of the critical region is not universal, the
quantitative implications for the chiral transition in QCD are uncertain.
%%%%%

%%%%%%%%%%%%%%%%%%%%%%%%%%%%%%%%%%%%%%%%%%%%

\section*{Acknowledgment}
We thank J\"{u}rgen Engels, Frithjof  Karsch and Bernd-Jochen Schaefer for
useful comments and stimulating discussions.
K.R. partial support from the
Polish Ministry of Science  and the
Deutsche Forschungsgemeinschaft (DFG) under the Mercator Programme.
B.S. gratefully acknowledges financial support from the
Helmholtz Research School on Quark Matter Studies.
%%%%%%%%%%%%%%%%%%%%%%%%%%%%%%%%%%%%%%%%%%%%%%%%

%%%%%%%%%%%%%%%%%%%%%%%%%%%%%%%%%%%%%%%
\appendix

%%%%%%%%%%%%   APPENDIX A     %%%%%%%%%%%%%%%%%%%%

%%%%%%%%%%%%%% APPENDIX B %%%%%%%%%%%%%%%%%%%%%%%%%%%%%%

%%%%%%%%%%%%%%%%%%%%%%%%%%%%%%

\end{document}